\documentclass[12pt]{iopart}
\usepackage{epsfig,mathrsfs,graphicx,iopams}

\def\be#1\ee{\begin{equation}#1\end{equation}}

\newcommand{\mean}[1]{\langle{#1}\rangle}
\newcommand{\ket}[1]{|{#1}\rangle}
\newcommand{\bra}[1]{\langle{#1}|}
\newcommand{\braket}[2]{\langle{#1}|{#2}\rangle}

\newcommand{\ba}{\begin{eqnarray} }
\newcommand{\ea}{\end{eqnarray} }

\begin{document}
\title{Noninvasiveness and time symmetry of weak measurements}
\author{Adam Bednorz$^1$, Kurt Franke$^{2,\footnotemark[1]}$, Wolfgang Belzig$^2$}
\address{$^1$ Faculty of Physics, University of Warsaw, Ho\.za 69, PL-00681 Warsaw, Poland}

\address{$^2$ Fachbereich Physik, Universit{\"a}t Konstanz, D-78457 Konstanz, Germany}
\footnotetext[1]{Present address: Max-Planck-Institut f{\"u}r Kernphysik und IMPRS-PTFS, Saupfercheckweg 1, 69117 Heidelberg, Germany}

\ead{Adam.Bednorz@fuw.edu.pl}

\begin{abstract}

Measurements in classical and quantum physics are described in fundamentally different ways.
Nevertheless, one can formally define similar measurement procedures with respect
to the disturbance they cause. 
Obviously, strong measurements, both classical and quantum, are invasive -- they
disturb the measured system. 
We show that it is possible to define general weak measurements, which are noninvasive:
the disturbance becomes negligible as the measurement strength goes to
zero. Classical intuition suggests that noninvasive measurements should be
time symmetric (if the system dynamics is reversible) and we confirm that
correlations are time-reversal symmetric in the classical case.
However, quantum weak measurements -- defined analogously to their classical
counterparts -- can be noninvasive but not time symmetric. We present a simple
example of measurements on a two-level system which violates time symmetry
and propose an experiment with quantum dots to measure the time-symmetry violation in a
third-order current correlation function.

\end{abstract}
\submitto{\NJP}
\maketitle

\section{Introduction}
The notion of a noninvasive measurement -- a measurement that does not disturb
the system being measured -- is undisputed in classical physics because one can assign
a real physical value to every point in phase space at all times. 
Even so, the situation becomes complicated if we introduce explicit detectors since
these may disturb the system.
In quantum physics, the notion of a noninvasive measurement is always problematic \cite{lega}. One cannot assign a value to an observable without discussing the measurement procedure. Strong projective measurements \cite{neumann} (and therefore the majority of general measurements \cite{povm,kraus}) are certainly invasive.

A good candidate for a noninvasive measurement scheme is a \emph{weak
measurement} \cite{aav}.
In general, by reducing the coupling of the detector system to the system
under measurement, the invasiveness is reduced at the price of an increased
detector noise.
This leads to paradoxes of unusually large values for single measurement
results after a subsequent postselection \cite{aav}, or a quasiprobability for the measured distribution after 
the detector noise has been removed\cite{ours}.
There is growing interest in such measurements \cite{wex1,korr,jor1,jor2,lgex}.

In this paper, we answer the question of when our intuitive criteria (defined
below) of noninvasiveness and time symmetry of measurements are
satisfied, for both classical and quantum cases.
Time-reversal symmetry of observables is a fundamental symmetry of physics, valid in
classical physics and in general -- because it is a good symmetry of quantum electrodynamics -- in
low-energy physics (in high-energy physics combined with parity and charge conjugation).
This symmetry is generally probed by the measurement of single,
non-time-resolved measurements, such as the measurement of electric dipole
moments of particles.
However, time-reversal symmetry also constrains the results of time-resolved
measurements with \emph{multiple} measurements.
For such considerations, one must consider the invasiveness of the
measurements themselves which will tend to break time-reversal symmetry.

\subsection{Measurement Schemes}

A \emph{measurement scheme} is a description of how to measure
\emph{observables}---functions of phase space for classical
physics or Hermitian operators for quantum physics.
A measurement takes place on a \emph{system under measurement} which is a member
of the \emph{ensemble under measurement}.
As usual, systems of the ensemble are considered to be identically distributed
and statistically independent.
Returning to the \emph{measurement scheme}, it should be a description of
(a) what the detector system
is and how it is prepared, (b) how the detector system is coupled to the
system under measurement, and (c) how the detector system is itself
measured, and how the measured value is interpreted.
The measurement scheme, essentially a description of the detectors, should be
generally independent of the \emph{ensemble under measurement}, and only (b),
the coupling to the system of interest, should depend on the observable.
Also, the measurement of the detector system must be defined in terms of
axioms---both classical and quantum (e.g. by projection postulate).
The measurement result should contain the \emph{inherent} statistical distribution $Q$ of the measured system.
The measurement result also contains \emph{detector noise} $D$
resulting, in an similar fashion, from the statistical and quantum properties of
the detector system.
By the measurement of many systems from an ensemble, the probability distribution
$P$ of the measurement can itself be measured.
The detector noise probability distribution $D$ of a \emph{null measurement}---a `measurement'
where the detector system is prepared but not coupled to the system under
measurement---can be determined.
We \emph{postulate} that the measurement scheme is expressed by a convolution, $P = Q
\ast D$ and in this case the detector noise may be removed by deconvolution.
The measurement schemes considered in this paper all possess this last
property.

\subsection{Noninvasiveness of measurements}

We consider time-resolved measurements of observables $A_1,\dots,A_n$
measured at times $t_1,\dots,t_n$, with outcomes $a_1,\dots,a_n$ occurring
with probabilities $Q(a_1,\dots,a_n)$.
The probability density $Q$ contains all the information about the experiment
and we formulate \emph{criteria} for noninvasiveness and time symmetry in
terms of $Q$, or more exactly, by requiring equality between $Q$ values measured in
different experiments.

An arbitrary operation is non-disturbing if the probability density of other measurements
is unchanged by the test operation's addition or removal.
In other words, integrating over the single measurement should yield the same
distribution that would be obtained if that measurement were never performed.
Therefore, our criterion of \emph{noninvasiveness} of the $k$th measurement reads
\begin{equation}
  \int \rmd a_k Q(a_1,\ldots,a_n)
  = Q(a_1, \ldots, \displaystyle{\not}a_k, \ldots, a_n).
  \label{eq:noninvasiveDefinition}
\end{equation}
Equation (\ref{eq:noninvasiveDefinition}) equates probabilities between two different experiments. In the
first, the $k$th measurement is integrated out and in the
second, the slash notation indicates that the variable was not
measured at all.
This defines noninvasiveness of single measurements on a given experiment.

More generally, if new measurements of observables $A_{k_1},\dots,A_{k_m}$ can be inserted at 
intermediate times without changing the previous probability density as in (\ref{eq:noninvasiveDefinition})
then the all of them are noninvasive.
The noninvasiveness is stronger if (\ref{eq:noninvasiveDefinition}) is satisfied for a fixed $A_k$ but arbitrary other 
measurements. 

\subsection{Time symmetry of measurements}

We assume that time reversal is a good symmetry for the
equations of motion of the system and investigate whether this leads to a corresponding symmetry expressed in the
results of measurements performed on the system.
We should note that time reversal symmetry holds only for nondissipative, Hamiltonian systems. 
However, physical dissipation is always a result of ignoring fast-changing and fine-grained degrees of freedom,
often modeled by a heat bath coupled weakly to the system. If one had access to all the degrees of freedom and
the heat bath, one could reverse the full phase space probability and restore
time symmetry. Even if it is not practically possible to reverse fine-grained
degrees of freedom, an alternate solution is to 
restrict ourselves to states in equilibrium coupled to a heat bath, which are time-symmetric themselves in the thermodynamic limit.

To express the expected time-reversal symmetry of a set of measurements,
we begin by denoting the time-reversed version of an object $X$ by $X^T$, i.e., position:
$q^T = q$ and momentum $p^T = -p$.
The time-reversed experiment involves the time reversed 
initial state $\rho\to\rho^T$, 
time-reversed measured quantities $A\to A^T$ with results $a\to a^T$, and also
reversed time---and therefore, ordering---of the measurements.
Hence, for the probability $Q$, our criterion of \emph{time symmetry of measurements} reads
\begin{equation}
  Q(a_1(t_1),\ldots,a_n(t_n))
  = Q^T(a^T_n(-t_n),\ldots,a^T_1(-t_1)),\label{tsym}
  \label{eq:timeSymmetryOfMeasurements}
\end{equation}
where we compare
the probability densities of the forward ($Q$)
and reversed ($Q^T$) sets of measurements.
In such a form, classically (\ref{tsym}) holds for
equilibrium and non-equilibrium systems and is
independent of the validity of charge conjugation and parity symmetries
and also of relativistic invariance \cite{cpt,cpviol, lor}.
When fulfilled---assuming for the moment that the measurements are
non-invasive---the result (\ref{eq:timeSymmetryOfMeasurements}) leads to the principle of
detailed balance\cite{kampen} and reciprocity of thermodynamic fluxes \cite{ons}.
\footnote{There exists also a different criterion of time symmetry, under
the exchange of boundary conditions in
pre- and postselected ensembles\cite{abl,ast}. It is satisfied even by invasive measurements,
so it is unrelated to ours.}

\subsection{Main result}

The above criteria (\ref{eq:noninvasiveDefinition}) and
(\ref{eq:timeSymmetryOfMeasurements}) must be confronted with real detection protocols.
For each measurement, there is a detection protocol that includes some interaction between the original system
and an ancilla that is later decoupled with the imprinted information retrieved from the system.
We should add the remark that the internal dynamics of the detector may be irreversible, but this is irrelevant, because we ask only about the 
behavior of the system. Note also that, for the time symmetry to hold,
the measurements should not disturb the
system in the sense of the criterion (\ref{eq:noninvasiveDefinition}), since any disturbance would create an asymmetry between before and
after the measurement.

The majority of measurements are invasive and irreversible, both classical and quantum.
However, there exists a special class of measurements, defined both classically and quantum mechanically, which are noninivasive under certain conditions.
They are described by an instantaneous interaction between the system and detector
$\sim gpA$, where $A$ is the measured observable, $p$ is the detector's momentum
and $g$ is the coupling strength (see details later in the text). 
The initial state of the detector is the zero mean Gaussian.
The observer finally registers the position which is shifted by $gA$.
The result contains also the internal detection noise, which is subtracted/deconvoluted.
For all \emph{finite} $g$ the scheme is invasive, except if the observables are compatible
(vanishing Poisson bracket or commutator) or if initially $p=0$, which makes sense only classically
(we do not want divergent position).

However, the scheme becomes noninvasive (both classically and quantum) \emph{in the limit} $g\to 0$,
while rescaling the detector's result  by $1/g$ -- this is the \emph{weak measurement} \cite{aav}.  
Surprisingly, classical and quantum weak measurements differ with respect
to time symmetry (\ref{tsym}). The behavior of different types of measurements is summarized in
Table \ref{tab_mes}.
The aim of this paper is to explain the origin of this difference between classical and quantum measurements.
We will also show the asymmetry explicitly by giving an example of a measurement of a simple two-level system
and propose an experimentally feasible realization by charge measurements on a quantum dot connected to a reservoir.
\begin{table}
\begin{tabular}{lll}
\hline
&Noninvasiveness & Time symmetry\\
\hline
General, strong&No&No\\
\hline
Classical $p=0$ (arbitrary $g$)&Yes&Yes\\
\hline
Compatible (arbitrary $g$) &Yes&Yes\\
\hline
Classical weak ($g\to 0$)&Yes&Yes\\
\hline
Incompatible quantum weak ($g\to 0$)&Yes&{\bf No}\\
\hline
Quantum weak ($g\to 0$) -- exceptions&Yes&Yes\\
\hline
\end{tabular}
\caption{Different types of measurements may satisfy noninvasiveness and/or time symmetry.  The exceptions include position and/or momentum measurement
in a simple harmonic oscillator, two-time correlations and other accidental symmetries or quasiclassical systems.}
\label{tab_mes}
\end{table}

\section{Time symmetry violation}

We will show in next sections that in the \emph{classical} weak measurement limit one can find
\begin{equation}
Q(a)=\langle\delta(a_n-A_n(t_n))\cdots\delta(a_1-A_1(t_1))\rangle,\label{cac}
\end{equation}
where the average $\langle \cdots\rangle=\int \rmd\Gamma \cdots\rho$ is taken in the initial state $\rho$ in the phase space $\Gamma$ and $A(t)$ denotes a classical analogue of the Heisenberg picture
for the observable $A$.
This clearly satisfies noninvasiveness and time symmetry, because $A$ are commuting numbers and we can reorder
them under time reversal.

Now, in the quantum case, we will get
\begin{equation}
Q(a)=\langle\delta(a_n-\check{A}_n(t_n))\cdots\delta(a_1-\check{A}_1(t_1))\rangle\label{qaq}
\end{equation}
for $t_n\geq\cdots\geq t_2\geq t_1$, where $\langle\cdots\rangle=\mathrm{Tr}\cdots\rho$ with the initial density matrix $\rho$. The superoperators act as $\check{A}B=(AB+BA)/2$, for the observable operator $A$.
This quantity is no longer a probability but a quasiprobability \cite{ours} and still satisfies noninvasiveness (\ref{eq:noninvasiveDefinition}).
However, the time symmetry
(\ref{eq:timeSymmetryOfMeasurements}) is violated, except for compatible measurements (e.g. space-like separated\cite{cpt}). 
Mathematically, this is because we replace the classical $c$-number multiplication
(obviously a commuting operation) by the quantum anticommutator of operators (therefore noncommuting).
We cannot reorder superoperators $\check{A}$ under time reversal.

For slow measurements, each operator $A(t)$ in (\ref{qaq})
is replaced with $\int f(t) A(t)\rmd t$,
where $f(t)$ turns on and off slowly compared to
relevant timescales of the system.
This slow measuring smoothes the resulting distribution $Q$ so that any
antisymmetric contributions vanish and therefore time symmetry
(\ref{eq:timeSymmetryOfMeasurements}) will still apply.
Roughly speaking, the more classical is the system, the more time-symmetric it is.

The time symmetry (\ref{eq:timeSymmetryOfMeasurements}) can be tested by comparing moments
of the distribution,
\begin{equation}
  \mean{a_1(t_1) \cdots a_n(t_n)}_Q
  = \mean{a^T_n(-t_n) \cdots a^T_1(-t_1)}^T_Q
  .
  \label{eq:t_moments}
\end{equation}
We emphasize that the quantities in (\ref{eq:t_moments}) are expectation
values of products of measurement results, and should not be confused with
expectation values of observables in an ensemble.
The ordering of $a_1$ to $a_n$ is mathematically irrelevant, but serves as a
reminder of the ordering of measurements in the experiment.

Linear correlations of quantum weak measurements---in the limit of zero measurement
strength---are given by
 \cite{ours,stjhcl}
\begin{equation}
\left\langle \prod_k a_k(t_k)\right\rangle_Q=\langle\check{A}_n(t_n)...\check{A}_2(t_2)\check{A}_1(t_1)\rangle
\label{tinor}
\end{equation}
We can freely permute the $a$ in the left hand side but not the $\check{A}$ in the right hand side (they do not commute and the order reflects that $t_n\geq\cdots\geq t_2\geq t_1$).
This asymmetry is only present for fast measurements of three or more
incompatible observables. This does not need a specific system. In contrast,
only specific systems and observables do not show the asymmetry; one such an exception is e.g. position measurement in
a simple harmonic oscillator.
In the case of compatible or only two (not necessarily compatible)
measurements the ordering is irrelevant and the symmetry (\ref{eq:t_moments})
holds.

\section{Direct measurements}

Let us take a classical system with the probability density $\rho(\Gamma)$ in phase space $\Gamma=(\Gamma_1,\dots,\Gamma_N)$ with $\Gamma_i=(q_i,p_i)$ being a pair of canonical generalized position and momentum.
The evolution is given by the Hamiltonian $H(\Gamma)$ and can be expressed
compactly using the Liouville operator $\check{L}$, defined by
$\check{L}A=(A,H)$ where 
\begin{equation}
(A,B)=\sum_i \left[
\frac{\partial A}{\partial q_i}\frac{\partial B}{\partial p_i}-\frac{\partial B}{\partial q_i}\frac{\partial A}{\partial p_i}\right]
\end{equation}
is the Poisson bracket. One has $\partial_t\rho=-\check{L}\rho$ or $\rho(t)=\rme^{-t\check{L}}\rho(0)$.

Let us consider a direct sequential measurement of quantities $A_1\dots A_n$ measured at times $t_1<t_2<\dots<t_n$, with the results $a_1\dots a_n$, respectively. 
The probability distribution is naturally postulated as
\begin{eqnarray}
&&Q(a)=\int \rmd\Gamma \delta(a_n-A_n)\rme^{(t_{n-1}-t_n)\check{L}}\cdots\nonumber\\
&&\delta(a_2-A_2)\rme^{(t_1-t_2)\check{L}}\delta(a_1-A_1)\rme^{-t_1\check{L}}\rho(0).\label{qdef}
\end{eqnarray}
Alternatively, it can be written as
\begin{equation}
Q(a)=\int \rmd\Gamma \delta(a_n-A_n(t_n))\cdots\delta(a_1-A_1(t_1))\rho(0)
\end{equation}
where $A(t)=\rme^{t\check{L}}A$.
The above quantity coincides with (\ref{cac}), is positive and normalized so it is a normal probability. 
As we already noted, it satisfies noninvasiveness (\ref{eq:noninvasiveDefinition})
and time symmetry (\ref{eq:timeSymmetryOfMeasurements}).

Now, the quantum direct measurement is governed by the projection postulate \cite{neumann}.
It is obviously invasive, violates (\ref{eq:noninvasiveDefinition}) and (\ref{eq:timeSymmetryOfMeasurements}), which is not at all surprising.
Looking for quantum noninvasiveness, we have to abandon direct measurements \cite{povm}.
Since we want to compare classical and quantum noninvasive measurements, we will
consider indirect measurements both classical and quantum.

\section{Weak measurements}

Let us now construct a model of a weak measurement which functions both
classically and quantum mechanically\cite{aav}. 
We have no direct access to the quantity $A$ at time $t_0$ but we
couple a detector for an instant.
The interaction Hamiltonian, added to the system, reads $H_I=g\delta(t-t_0)pA$
where $p$ is the detector's momentum, and $g$ is the measurement's strength.
We will use a very compact notation that highlights quantum-classical
analogies and differences. This is why many formulae below apply both to
classical and quantum cases, with differences only in the mathematical objects
(e.g. numbers or operators, phase-space density or density matrix, operator
or superoperator).
The quantum Liouville superoperator reads $\check{L}A=[A,H]/\rmi\hbar$
(commutator $[A,B]=AB-BA$). As in the classical evolution of phase space
density, operators in the Heisenberg picture evolve as $A(t)=\rme^{t\check{L}}A$.
For a single measurement, the total initial state is a product
$\rho_\rmd\rho(t_0)$, where $\rho_\rmd$ is the state of the detector. After the
measurement, the total density is
\begin{equation}
\rho_\rmd\rho(t_0)\to\exp(g(\check{A}\check{p}^q+\check{p}\check{A}^q))\rho_\rmd\rho(t_0),
\end{equation}
where classically $\check{A}=A$ (multiplication by $A$), quantum-mechanically
$\check{A}B=\{A,B\}/2$ (anticommutator $\{A,B\}=AB+BA$), and the (super)operator $\check{A}^q$ is given classically by $\check{A}^qB=(A,B)$,
and quantum mechanically by $\check{A}^qB=[A,B]/\rmi\hbar$. For a more conventional approach, see Appendix A.
Note that classically $\check{p}^q=-\partial_q$ for the canonically conjugated $q$. The analogy between classical Poisson brackets and quantum commutators
was recognized in the early days of quantum mechanics \cite{dirac}. The novel analogy here is between the classical
multiplication by an observable and the quantum anticommutator. The fact that
we replace a (commuting) number by a (noncommuting) superoperator helps to
understand why quantum weak measurements do not obey time symmetry while
classical weak measurements do.

\begin{figure}
\includegraphics[scale=.3]{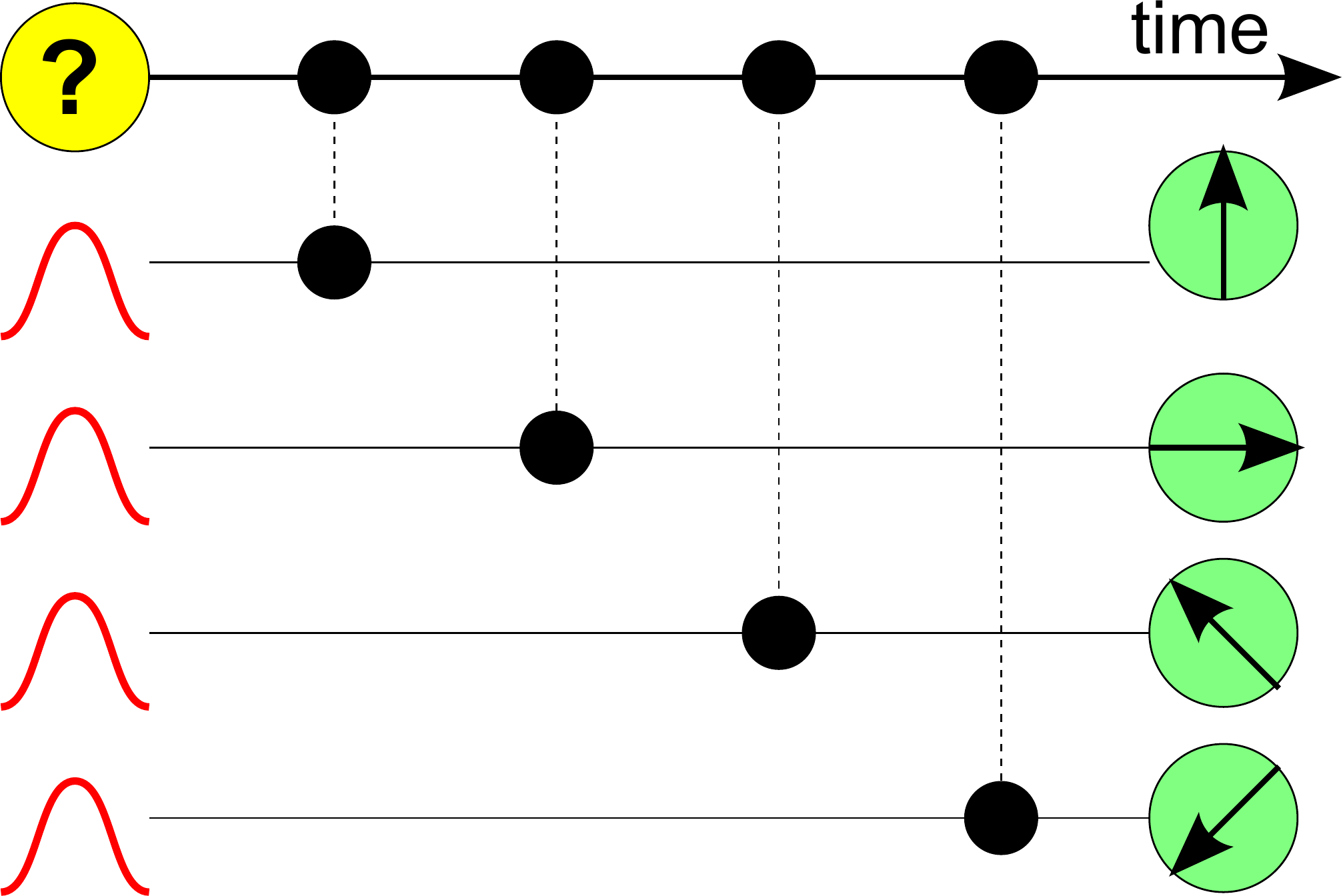}
\caption{A schematic diagram of weak measurements, analogous to figure 3 in \cite{jor1}. The measured system (yellow) instantly interacts with a prepared ancilla 
(red state),
which is measured (projectively in the quantum case) afterwards (green detector). The procedure is repeated with identical but independent ancillae.}
\label{weakm}
\end{figure}

If we discard the results of the measurement then the resulting density reads 
$\langle \exp(gp\check{A}^q)\rho\rangle$, where the average denotes $\int \rmd\Gamma_\rmd\cdots\rho_\rmd$ classically
and $\mathrm{Tr}_\rmd\cdots\rho_\rmd$ quantum mechanically (subscript $\rmd$ denotes detector's subspace).
The procedure can be repeated for sequential measurements as depicted in  figure \ref{weakm}.

We take the initial state of the detector given by
\begin{equation}
\rho_\rmd\propto \exp(-q^2/2\alpha-p^2/2\beta)
\end{equation}
where $q,p$ are a pair of conjugate canonical observables (with the property $(q,p)=1$ or $[q,p]=\rmi\hbar$). 
This is a generic symmetric Gaussian state. If measured classically the initial variances read
$\langle q^2\rangle\equiv\sigma_q=\alpha$ and $\langle p^2\rangle\equiv\sigma_p=\beta$.
Quantum mechanically (under projective measurement) $\sigma_q=(\hbar/2)\sqrt{\alpha/\beta}\coth\sqrt{\hbar^2/4\alpha\beta}$
and $\sigma_p=(\hbar/2)\sqrt{\beta/\alpha}\coth\sqrt{\hbar^2/4\alpha\beta}$.
Note that for $\alpha\beta\gg \hbar^2$ they reduce to the classical result while $\sigma_q\sigma_p\geq\hbar^2/4$
is imposed by the Heisenberg uncertainty principle.

We register directly the value of $q$.
However, the way of measuring of $q$ is in principle irrelevant, both classical and quantum, and may be well disturbing
because the detector will not interact with the system anymore.
The detector (classical or quantum) can evolve irreversibly, we are only interested in the data
extracted from the system.

We apply a sequence of such measurements, using identical, independent detectors $q_1,\dots,q_n$, but coupled at different times
to possibly different observables. 
It is convenient to define a \emph{result-conditioned} density $\rho_g(q)$, normalized by the final 
\emph{result-integrated} density $\rho_g=\int \rmd^nq\; \rho_g(q)$. The probability density of a given sequence of results is given by
$P(q)=\int \rmd\Gamma\rho_g(q)$ or $\mathrm{Tr}\rho_g(q)$. Now, $\rho_g(a)$ is given by
\begin{equation}
\rho_g(q)=\int \rmd^na\;\varrho_g(a)\prod_k D(q_k-ga_k),\label{roconv}
\end{equation}
where $D$ is the zero-mean Gaussian noise with the variance $\sigma_q$.
The quantity $\varrho_g(a)$  reads 
\begin{eqnarray}
&\varrho_g(a)=&\rme^{\sigma_p(g\check{A}_n^q)^2/2}\delta(a_n-\check{A}_n)\rme^{(t_{n-1}-t_n)\check{L}}\cdots\nonumber\\
&&\times\rme^{\sigma_p(g\check{A}_2^q)^2/2}\delta(a_2-\check{A}_2)\rme^{(t_1-t_2)\check{L}}\rme^{\sigma_p(g\check{A}_1^q)^2/2}\delta(a_1-\check{A}_1)\rme^{-t_1\check{L}}\rho(0).\label{qweak}
\end{eqnarray}
This is classically a standard probability density but not a positive definite density matrix in quantum mechanics.
It is clear when defining $Q_g(a)=\int \rmd\Gamma\varrho_g(a)$ and $Q_g(a)=\mathrm{Tr}\varrho_g(a)$.
Now the quantum $Q_g$ is only a quasiprobability \cite{ours}.
One can write down the convolution relation analogous to (\ref{roconv}),
\begin{equation}
P(q)=\int \rmd^na\;Q_g(a)\prod_k D(q_k-ga_k).
\end{equation}
Both $\varrho_g$ and $Q_g$ have a well-defined limit $g\to 0$, $\varrho\equiv\varrho_0$ and $Q\equiv Q_0$. Then (\ref{qweak}) reduces
to (\ref{qdef}) classically. In the quantum case,
\begin{equation}
Q(a)=\mathrm{Tr} \delta(a_n-\check{A}_n(t_n))\cdots\delta(a_1-\check{A}_1(t_1))\rho(0)\label{qweakq0}
\end{equation}
with $\check{A}(t)B=\{A(t),B\}/2$ or equivalently $\check{A}(t)=\rme^{t\check{L}}\check{A}\rme^{-t\check{L}}$,
which coincides with (\ref{qaq}). The effect of disturbance (both classical and quantum!) is of the order $g^2$ so it vanishes in the limit $g\to 0$.

One can relate correlation functions
\begin{equation}
\langle q_1\cdots q_n\rangle_P=g^n\langle a_1\cdots a_n\rangle_Q.
\end{equation}
The leading contribution to such correlation functions is of the order $g^n$, while
the lowest correction due to disturbance is of the order $g^{n+2}$, as follows
from (\ref{roconv}) and (\ref{qweak}).

Both classical and quantum $Q$ satisfy noninvasiveness (\ref{eq:noninvasiveDefinition}), but only in the $g\to 0$ limit. There are exceptions
when noninvasiveness holds for an arbitrary $g$. In particular $Q_g=Q$ is independent of $g$ and always a real positive probability for \emph{compatible} observables -- if $(A_j(t_j),A_k(t_k))=0$ classically 
or $[A_j(t_j),A_k(t_k)]=0$ quantum mechanically for all $j,k$.
We emphasize that the deconvolved result-conditioned density $\varrho(a)$ (\ref{qweak})
changes with each measurement
because it gets the factor $\delta(a-A)$ or $\delta(a-\check{A})$.  This is because it
must contain the read-off knowledge (it is gaining information -- not disturbance). It is impossible to preserve
the result-conditioned density unchanged by any measurement, both classical or quantum (unless the measurement is
void) -- in this sense all measurements would be invasive. Hence, only after integration it makes sense to 
distinguish between invasive and noninvasive measurements.

From (\ref{roconv}) and (\ref{qweak}) we see also that the result-integrated density after a single weak measurement gets the factor $\rme^{\sigma_p(g\check{A}^q)^2/2}$, which reduces to identity in the limit $g\to 0$. This is why 
weak measurements (both classical and quantum) are noninvasive in a stronger
sense: their disturbance vanishes as $g^2$ regardless of the type of 
measurements before/after. For a comparison, strong measurements of compatible observables are mutually noninvasive
but we can find an incompatible observable whose results they disturb.
The price of weak measurements is that one has to repeat the experiment $\gtrsim 1/g^2$ times to get the weak signal out of 
statistics.

Note also that the scaling $q\sim ga$ is analogous in the classical and quantum cas.
In the classical case, however, one can take $\sigma_p=0$, which makes the limit $g\to 0$ unnecessary.
On the other hand, the quantum mechanical uncertainty principle allows only for the limiting noninvasiveness.
One could argue (both classically and quantum) that there is still some invasiveness for large results because
the result-conditioned density $\rho_g(q)$ is affected by different factors for different values of $A$.
Namely, $\exp(-(q-gA)^2/2\sigma_q)/\exp(-(q-gA')^2/2\sigma_q)$ can be large even for small $g$.
However, this requires $q\gtrsim q_0=\sigma_q/gA$ which happens very rarely for small $g$, with the estimated probability
of the rapidly vanishing Gaussian tail $\sim \rme^{-q_0^2/2\sigma_q}=\rme^{-\sigma_q/2(gA)^2}$ so it is irrelevant
for the discussion of noninvasiveness. Moreover, $\rho_g(q)$ also contains the read-off knowledge, although 
rescaled by $g$, while only the change of result-integrated $\rho_g$ is quantifies invasiveness.

\subsection{Causality}

One may ask whether it is possible to enforce time-symmetry (\ref{eq:timeSymmetryOfMeasurements}) in any other measurements scheme.
Unfortunately, we would pay a high price -- abandoning causality of measurements.

All general quantum measurements appear in a \emph{causal} way,
\begin{equation}
P(q)=\mathrm{Tr}\check{K}_n(q_n)\rme^{(t_{n-1}-t_n)\check{L}}\cdots\check{K}_1(q_1)\rme^{-t_1\check{L}}\rho(0)
       \end{equation}
with normalized completely positive maps $\check{K}$ \cite{povm,kraus}.
Even more generally
\be
P(q)=\mathrm{Tr}\mathcal T\check{K}[A,q]\rho,
\ee
where $\mathcal T$ denotes time ordering of superoperators that depend on observables $A(t)$ in Heisenberg picture.
Now, every causal measurement of non-zero strength is disturbing (weak measurements from section 3 create a disturbance $\sim g^2$) but only \emph{forward} in time. 
If we measure at $t_1<t_2<t_3$ then the measurement $1$ disturbs $1,2,3$, the measurement $2$ disturbs only $2,3$
and the last disturbs only itself. If there existed any measurement scheme with the time-symmetric limit 
(with a vanishing parameter analogous to $g$)
then it would have also time-symmetric disturbance at finite strength -- violating causality.

However, if we give up the above rule or are satisfied by only limiting causality (at $g=0$) we can e.g. define
\be
Q(a)=\int\frac{\rmd^n\chi}{(2\pi)^n}\mathrm{Tr}\exp\sum_k\rmi(a_k-A_k(t_k))\chi_k\:\rho.
\ee
The corresponding map $\check{K}B=KBK^\dag$  for $n$ measurements reads
\be
K(q)=(2\pi)^{-n/4}
\rme^{\sum_k (2gA_k(t_k)q_k-q_k^2)/4}\left[\mathcal T_s\rme^{-\left(\sum_k gA_k(t_k)\right)^2/2}\right]^{1/2},
\ee
where 
$\mathcal T_s$ denotes the rule of complete symmetrization of operator products in Taylor expansion.
The probability $P=\mathrm{Tr}\check{K}\rho$ is related to $Q$ by (\ref{roconv}) with $\sigma_q=1$.
It is perfectly time-symmetric but the disturbance is time symmetric, too, for $g>0$.
In this work, we have not considered this option, because all known experimental detection schemes confirm 
causality.

\section{Examples}

\subsection{Double well}

\begin{figure}
\includegraphics[scale=.7]{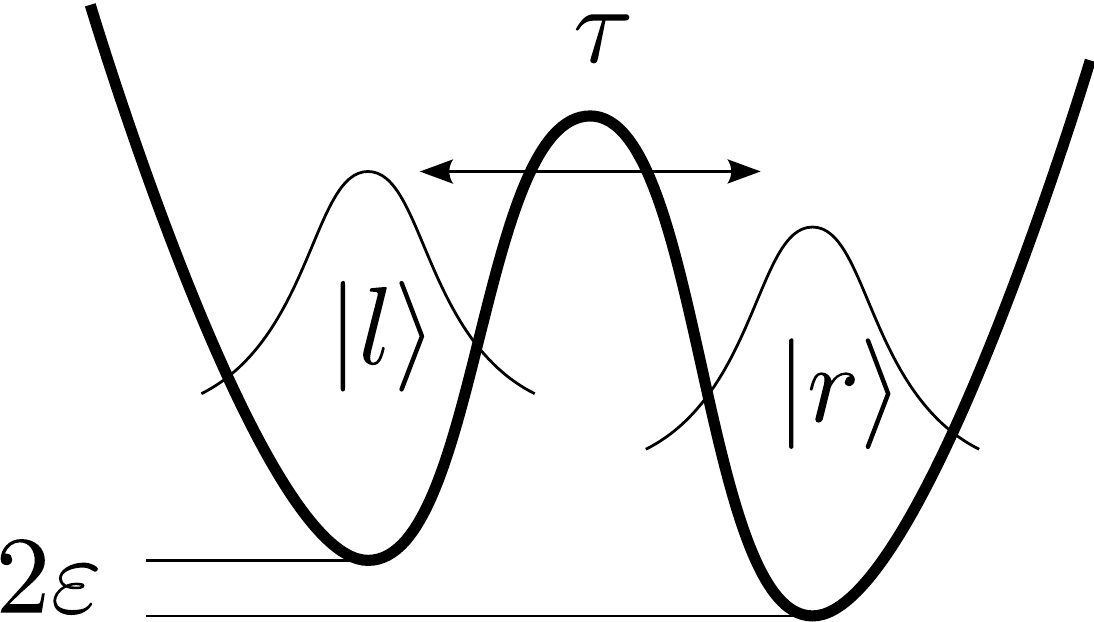}
\caption{The double well, described effectively by two states $\ket{l}$ and $\ket{r}$,
with energy shift $2\varepsilon$ and tunneling amplitude $\tau$. 
Even in the ground state, quantum fluctuations allow jumps between the wells, which turn to be non-time symmetric under weak measurement.
}\label{dwell}
\end{figure}

Let us demonstrate the paradox in a simple system consisting of a particle
in a double-well potential as in figure\ \ref{dwell}. For simplicity, we take an equilibrium
state, but the asymmetry appears also in a completely general case.
The particle is effectively described by the ground states of the left and
right wells, $\ket{l}$ and $\ket{r}$ respectively.
Higher excited states have much more energy and for low temperatures
can be ignored, leaving an effective two-state system.
Using the basis states,
the operator for expected location is $Z= \ket{l}\bra{l}-\ket{r}\bra{r}$, and
the effective Hamiltonian reads
\begin{equation}
  H =\varepsilon(\ket{l}\bra{l}-\ket{r}\bra{r})+\tau(\ket{l}\bra{r}+\ket{r}\bra{l}),
\end{equation}
where $2\varepsilon$ is the energy difference between wells and $\tau$ is the tunneling amplitude.

For low-energy physics, time reversal alone is already a good symmetry
in the equations of motion so, in absence external
magnetic field, the equilibrium state is time symmetric.
Hence, $H$ and $Z$ are even under time reversal ($H^T=H$, $Z^T=Z$).
We are now in a position to test equation (\ref{eq:t_moments})
with $z$ measured at three separate
times and with the initial thermal state $\rho\propto\exp(-H/kT)$.
The correlation for three weak measurements can be calculated using
(\ref{tinor}) and $Z(t)=\rme^{\rmi Ht/\hbar}Z\rme^{-\rmi Ht/\hbar}$:
\begin{equation}
  \langle{z(t_1)z(t_2)z(t_3)}\rangle=\alpha
    (\varepsilon^2 + \tau^2 \cos(2(t_3-t_2) \Delta/\hbar)),
  \label{quantum_dot_result}
\end{equation}
where $\Delta = \sqrt{\varepsilon^2 + \tau^2}$, $\alpha=-(\varepsilon/\Delta^3)\tanh(\Delta/kT)$.
For this system and measurements, the expression corresponding to the right-hand side of (\ref{eq:t_moments}) differs from (\ref{quantum_dot_result})
only by the exchange of $t_3-t_2$ with $t_2-t_1$.
However, (\ref{quantum_dot_result}) is clearly asymmetric under
this exchange,
demonstrating that time-reversal symmetry
is broken for correlations of quantum weak measurements.
As a side note, it can be shown that the correlation
(\ref{quantum_dot_result}) is independent of
measurement strength; however, this coincidence does not hold in general.

\subsection{Quantum dot}

Despite the simplicity of the above example, a genuine, fast weak detection
scheme is probably difficult to implement experimentally in this case.
Below, we present a more realistic example, leveraging recent
developments in quantum dots \cite{dotab}.
\begin{figure}
\begin{minipage}[b]{0.3\linewidth}
\includegraphics[scale=.5,angle=0]{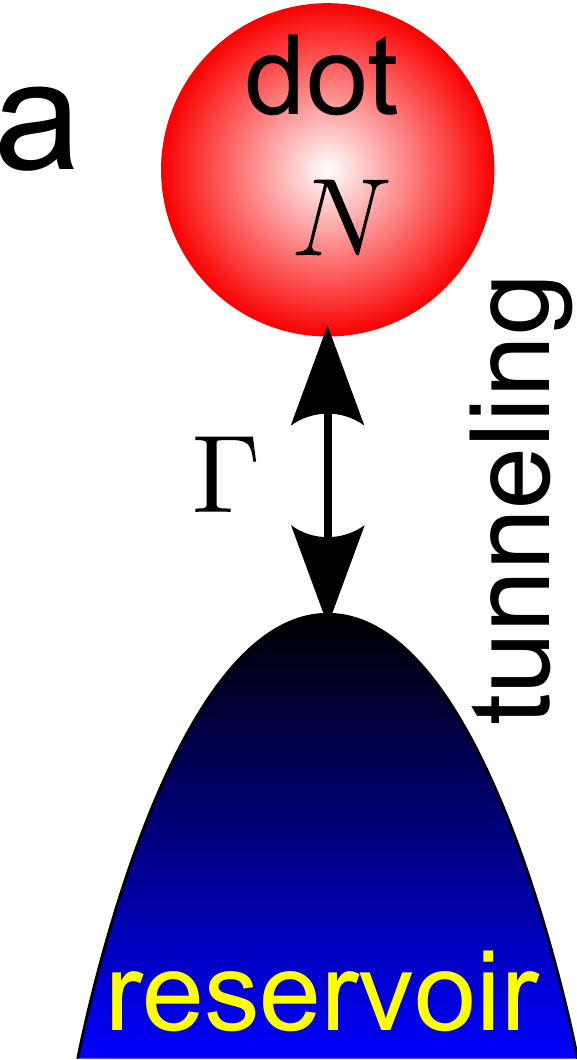}
\vspace{0.5cm}

\includegraphics[scale=.5,angle=0]{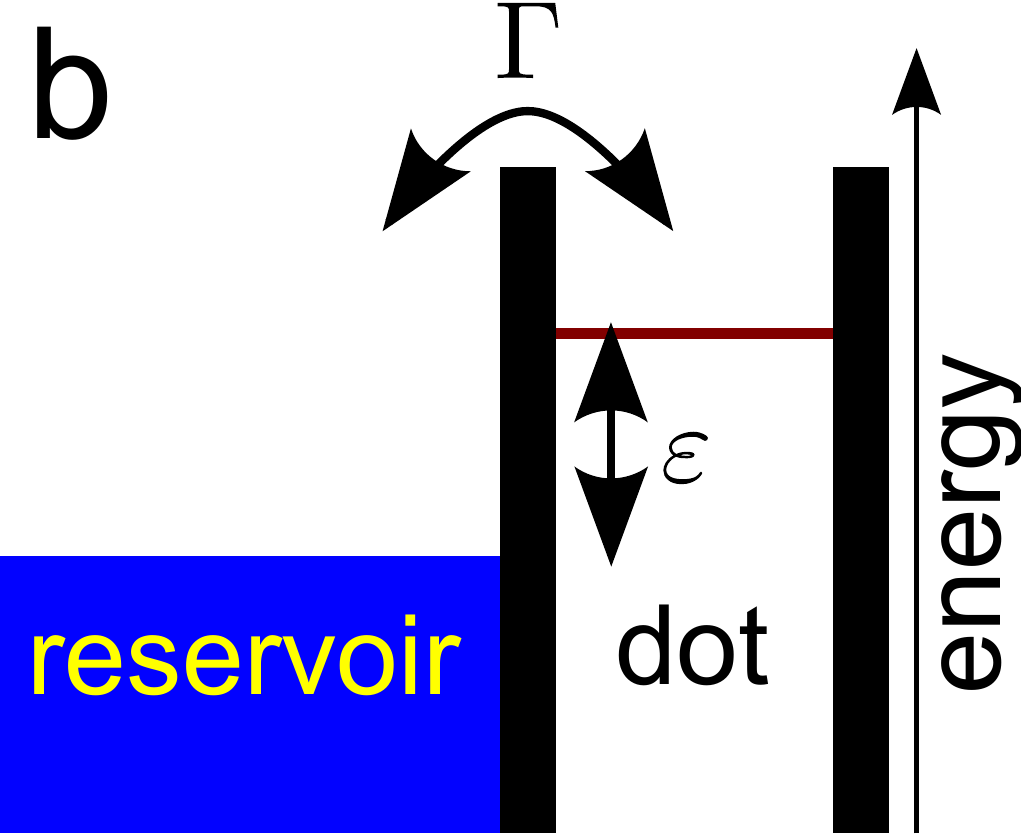}
\end{minipage}
\begin{minipage}[b]{0.3\linewidth}
\includegraphics[scale=.5,angle=0]{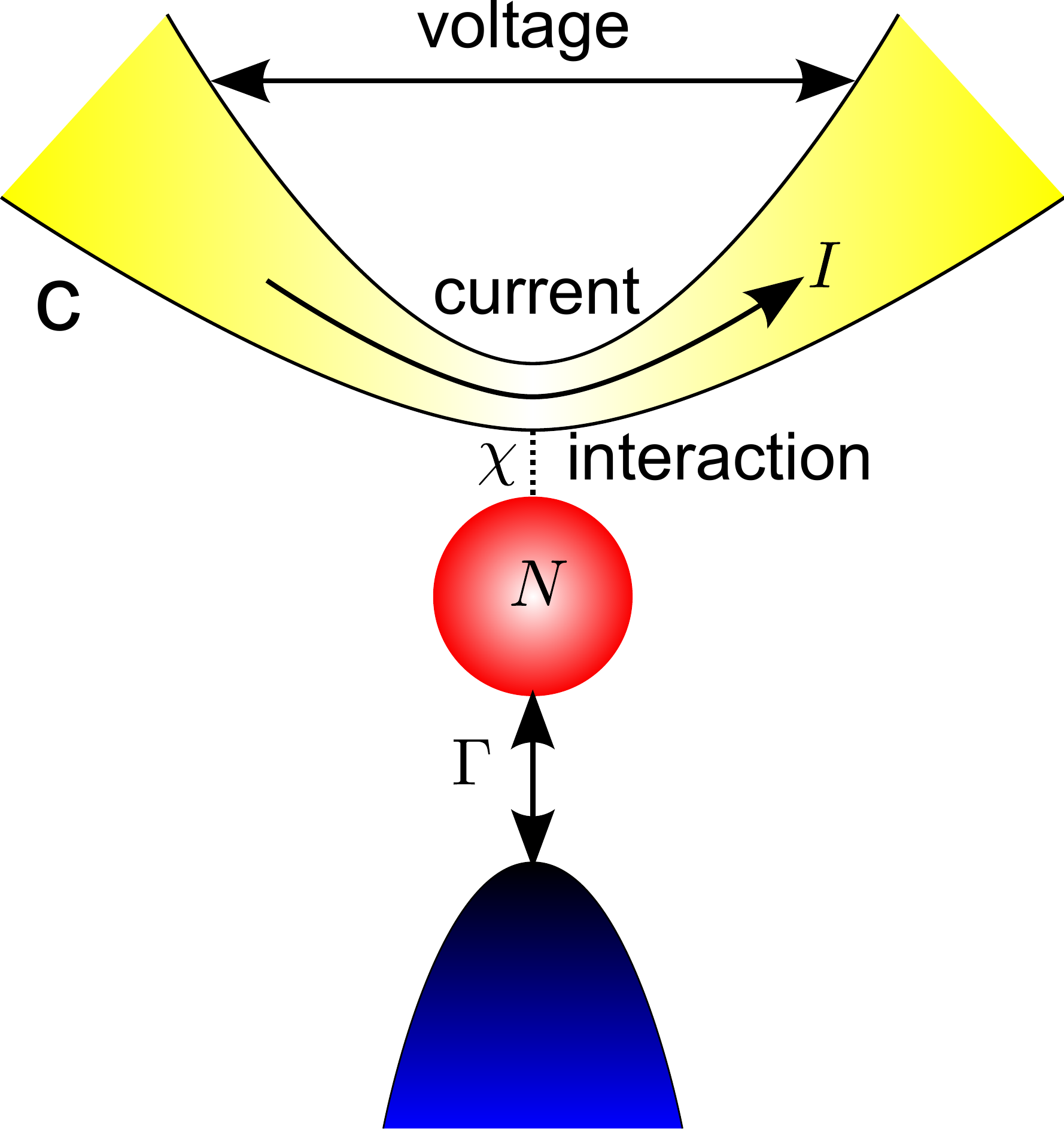}
\end{minipage}
\caption{(a) The system consists of the dot (red) exchanging the elementary charge with a reservoir (blue).
(b) The diagram of energy levels. At zero temperature, the Fermi sea of blue levels is full
but electrons may still jump on and off the dot's red level.(c) Proposed detection by an electric junction (yellow).
The junction and dot are weakly coupled capacitively and the directly measured quantity is the current $I$ through the junction biased with the voltage. The charge is allowed to jump between the dot and the lower reservoir but not the junction.
}\label{doo}
\end{figure}
We consider a quantum dot containing a single energy level $\varepsilon$, coupled to a Fermi reservoir by
an energy independent coupling
described effectively by the tunneling rate $\Gamma/\hbar$, as depicted in figures \ref{doo}(a) and (b).
The occupation $n$ on the dot (classically either $0$ or $1$ in elementary charge units)  is the measured
observable $N$.
The quantum observable and the Hamiltonian read \cite{blanter}
\begin{equation}
N=c^\dag c,\:H=\varepsilon N+\label{hdot}
\int \rmd E\; [\sqrt{\Gamma/2\pi} c^{\dag}\psi(E)
+\mbox{h.c.}+E\psi^{\dag}(E)\psi(E)],\nonumber
\end{equation}
which
describes energy-independent tunneling between the dot and reservoir, where
$\varepsilon$ is the dot level energy.
We assume usual fermion anticommutation relations $\{\psi,\phi\}=0$, $\{\psi^\dag,\phi\}=0$ if $\psi\neq \phi$, $\{c^{\dag},c\}=1$
and $\{\psi^{\dag}(E),\psi(E')\}=\delta(E-E')$.
Spin is neglected here but if necessary all results can be simply multiplied by $2$. 
The initial state is $\rho\propto\exp{(-H/kT)}$.
The Hamiltonian (\ref{hdot}) and the occupation are certainly symmetric under time reversal, $H^T=H$ and
$N^T=N$.

\begin{figure}
\includegraphics[scale=1.2,angle=0]{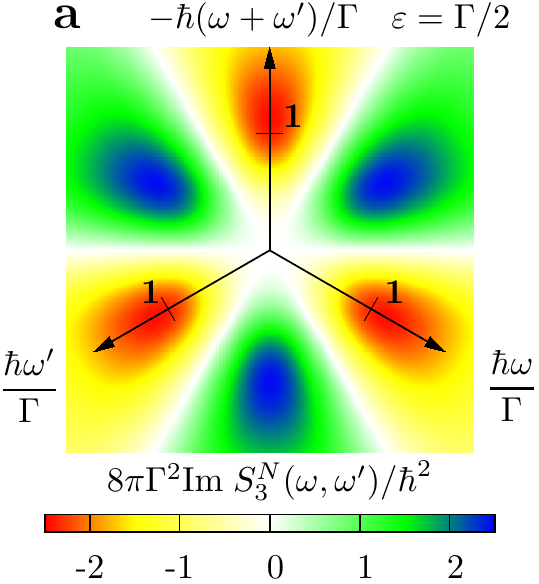}
\includegraphics[scale=1.2,angle=0]{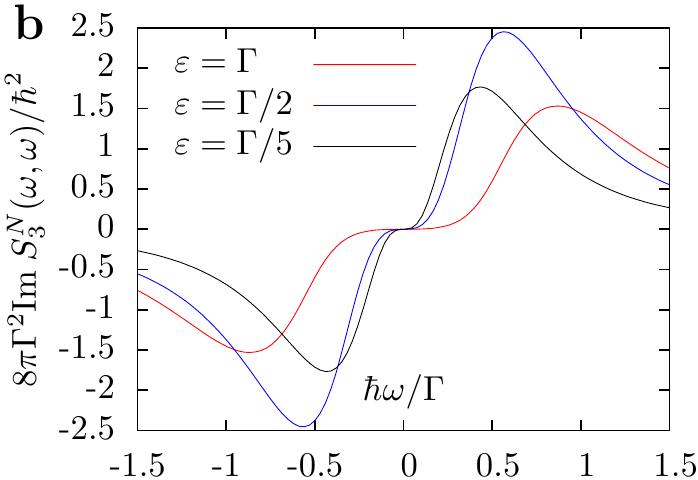}
\caption{Time asymmetry of the third-order correlation of dot occupation fluctuations $S_3^N(\omega,\omega')$ in frequency domain at $kT\ll\Gamma$. The symmetry is broken if $\mathrm{Im}\;S_3^N\neq 0$. (a) For $\varepsilon/\Gamma=0.5$ and arbitrary $\omega,\omega'$
 the asymmetry vanishes (white lines) at when $\omega$, $\omega'$ or $\omega+\omega'$
is equal to $0$. The hexagonal structure reflects the symmetry under permutations of frequencies.
(b) Taking different values
$\varepsilon/\Gamma=1,0.5,0.2$ and $\omega=\omega'$ it is clear that the maximal asymmetry occurs for  energy/frequency parameters of the same order.}
\label{assym}
\end{figure}

To show the time asymmetry we will use the frequency domain, defining the third cumulant
\begin{equation}
S^N_3(\omega,\omega')=\int \rmd t\rmd t'\;\rme^{\rmi\omega t+\rmi\omega't'}
\langle \delta n(t)\delta n(t')\delta n(0)\rangle\label{tri}
\end{equation}
with $\delta n=n-\langle n\rangle$.
The asymmetry-probing quantity is the imaginary part of the third cumulant $\mathrm{Im} S^N_3(\omega,\omega')$, which should vanish
if (\ref{eq:t_moments}) holds.
To calculate (\ref{tri})  we use the 
close-time-path formalism \cite{ctp,utsumi:07},
defining matrices in $2\times 2$ Keldysh space
\begin{equation}
\check{N}=\left(\begin{array}{cc}
1&0\\
0&1/4\end{array}\right),\:
\check{G}=\left(\begin{array}{cc}
G^K&G^R\\
G^A&0\end{array}\right)
\end{equation}
with $G^R(\omega)=\rmi\hbar/(\hbar\omega-\varepsilon+\rmi\Gamma/2)=-G^{A\ast}(\omega)$
and $G^K(\omega)=\tanh(\hbar\omega/2kT)
\hbar\Gamma/(2(\hbar\omega-\varepsilon)^2+\Gamma^2/2)$. Then
\begin{equation}
S^N_3(\omega,\omega')=-\int\frac{\rmd\alpha}{2\pi}\mathrm{Tr}
\check{G}(\alpha)\check{N}[\check{G}(\alpha+\omega)+\check{G}(\alpha+\omega')]
\check{N}\check{G}(\alpha+\omega+\omega')\nonumber
\check{N}.
\end{equation}
The integral can be performed analytically but the result contains digamma functions at finite temperatures.  As
suspected, $\mathrm{Im} S^N_3$ is not zero, see figure \ref{assym}. Both imaginary and real parts vanish far from resonance.
The asymmetry is the strongest at low temperatures ($kT\ll \varepsilon$) and for
comparable energy, tunneling and frequency scales
($\varepsilon\sim \Gamma\sim\hbar\omega$).
This suggests that zero-point fluctuations of the charge jumping on and off the dot are responsible for the asymmetry.
The symmetry is restored if one of $\omega$, $\omega'$, or $\omega+\omega'$ is equal to $0$.
As expected, $\mathrm{Im} S^N_3$ vanishes for slow measurements
$\omega,\omega' \ll \Gamma/\hbar$.
In the limit $\omega,\omega'\to 0$, the result for $S^N_3$ is a special case of
the application of full counting statistics \cite{utsumi:07}.

For the experimental confirmation of the asymmetry one must introduce a weakly coupled detector.
We propose an electric voltage-biased junction coupled weakly to the dot, so that its conductance depends on the charge on the dot, see figure \ref{doo}(c).
The externally measured quantity is  the current $I$ through the junction, in particular
$I\simeq I^0+\chi N$, where $I^0$ is the intrinsic current in the junction and $\chi$ is its susceptibility due to the dot's charge. Then
$S_3^I\simeq S_3^{I0}+\chi^3S_3^N$
where $S_3^{I0}$ is the internal current noise of the junction in the absence of the dot.
Measurements of $S_3^{I0}$ have been demonstrated
\cite{reulet-reznikov}; therefore we
expect the measurement of $S_3^{I}$ to be feasible. Such an experiment will confirm the 
time-reversal symmetry violation only if the dot is not driven out of equilibrium.
It is always possible for a certain parameter range---see Appendix B for the detailed model.
Note that the dynamics of the detector here is clearly irreversible as it is initially 
in a nonequilibrium stationary state. However, we are only interested in the behavior of the system. Anyway, in the range of frequencies of possible asymmetry,
 $S^{I0}$ is frequency-independent, so the asymmetry of $S^I$ will show asymmetry of $S^N$.

\section{Conclusions}
We have shown that neither noninvasiveness  (\ref{eq:noninvasiveDefinition})
nor time symmetry (\ref{eq:timeSymmetryOfMeasurements}) is automatically 
satisfied in the results of measurements, both classical and quantum.
Only a subclass of detection schemes, parameterized by the measurement strength $g$, may satisfy (\ref{eq:noninvasiveDefinition}) and/or (\ref{eq:timeSymmetryOfMeasurements}).
Classically, the measurement can be strictly noninvasive either at a finite $g$ and zero detector's momentum
or in the weak measurement limit, $g\to 0$.
However, quantum  noninvasiveness is satisfied only in a the limit of zero strength $g\to 0$.
Moreover, the time symmetry  of measurements (\ref{eq:timeSymmetryOfMeasurements}) is broken in the quantum case, in contrast
to classical mechanics.
This is the fundamental difference between classical and quantum noninvasive measurements.
One could argue that the weak measurement still affects the system and forces
a time direction in this way.
On the other hand, one expects a natural limit in which the influence on the system is negligible and the time symmetry should hold.

This violation is effectively a failure of weak measurements to accurately reflect
the time-reversal symmetry inherent in a system.
 As such, it is independent of the validity of other symmetries such as
charge parity time.
Since quantum measurements of finite strength manifestly break time-reversal
invariance, our result shows that, in contrast to classical measurements, all
quantum measurements break time-reversal invariance regardless of their
strength. Weak measurements are then still disturbing in some sense, although they
do not disturb the state or later measurements.

Our result shows not only the quantum violation of time symmetry, but also the importance
of a classical-quantum analogy of detection schemes. An open question is to what extent the analogy is correct.
For instance, maybe not all system-detector interactions are allowed and possibly they cannot be
instantaneous but rather time-extended. This needs further research, referring also to
realistic experimental detection schemes.

\section*{Acknowledgements}
 
We are grateful to Bertand Reulet, Christoph Bruder
and Tomasz Dietl for helpful discussions.
We acknowledge financial support from the DFG
through SFB 767 and SP 1285 and by the Polish Ministry of Science grant IP2011  002371 (to AB).

\section*{Appendix A} 
\renewcommand{\theequation}{A.\arabic{equation}}
\setcounter{equation}{0}

To justify (\ref{tinor}) we consider a series of weak measurements.
Following Aharonov \emph{et al.} \cite{aav},
each weak measurement introduces an ancilla system and creates entanglement
via an instant interaction Hamiltonian $\hat{H}_I=\hbar\delta(t)g\hat{p}\hat{A}$ where $g$ is the strength of interaction,
$\hat{p}$ is momentum operator of the ancilla, conjugate to position $\hat{q}$ ($[\hat{q},\hat{p}]=i\hbar$), and $\hat{A}$ is the
measured observable.  The interaction is followed by von Neumann projection \cite{neumann} of
the ancilla onto a position eigenstate which
destroys the ancilla. The system can however be measured again with the next
ancilla, as shown in figure\ \ref{weakm}. The density matrix after the
$j$\textsuperscript{th} measurement is
\begin{equation}
  \hat\rho_j =
    \rme^{-\rmi g_j\hat{p}_j \hat{A}_j/\hbar}
    \left(
      \hat\rho_{j-1} \otimes
      \ket{\phi_j} \bra{ \phi_j}
    \right)
    \rme^{\rmi g_j\hat{p}_j \hat{A}_j/\hbar},
\end{equation}
where $\ket{\phi_j}$ is the initial prepared state of ancilla $j$.
By inserting identity operations
$\int da\, \ket{a}\bra{a} = \hat{1}$,
the measurement interaction can be expressed
as shifts of the ancilla wavefunction,
\begin{equation}
  \hat\rho_j =
    \int \rmd a'_j \, \rmd a''_j
        \ket{ \phi_j(q_j-g_ja'_j) }
        \bra{ \phi_j(q_j-g_ja''_j)}
             \ket{a'_j}\bra{a'_j}{\hat\rho}_{j-1}\ket{a''_j}\bra{a''_j}.
  \label{eq:weak_rho_recursive_with_idents}
\end{equation}
In (\ref{eq:weak_rho_recursive_with_idents}), the 
the state of ancilla $j$ which has the shifted
wavefunction $\phi_j(x_j-g_ja'_j)$
is written as $\ket{ \phi_j(x_j-g_ja'_j) }$.
The joint probability $P(q_1,\ldots,q_n) =: P(\boldsymbol{q})$
is the probability of measuring the ancillas in a set of
position eigenstates with positions given by $q_k$
\begin{eqnarray}
  P(\boldsymbol{q}) &=&
    \mathrm{Tr}
      \left\{
        \hat\rho_n
        \prod_k
          \ket{q_k}\bra{q_k} 
      \right\}\label{eq:joint_prob_with_integrals}\\
&=&\int \rmd\boldsymbol{a'} \, \rmd\boldsymbol{a''}\,\delta(a'_n-a''_n)
        \tilde{\rho}_n \left(\boldsymbol{a'},\boldsymbol{a''}\right) 
      \prod_k
      \phi\left(q_k-g_ka'_k)\right)
      \phi^\ast\left(q_k-g_ka''_k)\right).
      \nonumber
\end{eqnarray}
In (\ref{eq:joint_prob_with_integrals}),
$\tilde{\rho}_j$ is defined recursively by
\begin{equation}
  \tilde{\rho}_j (a'_1,a''_1,\ldots,a'_j,a''_j)
  =
    \braket{a'_j}{a'_{j-1}}
    \tilde{\rho}_{j-1} 
    \braket{a''_{j-1}}{a''_j} .
\end{equation}
Using Gaussian wavefunctions $\phi(q) = (2\pi)^{-1/4}\, \rme^{-q^2/4}$,
a change of variables to
$\boldsymbol{\bar a} = (\boldsymbol{a'}+\boldsymbol{a''})/2$
and $\boldsymbol{\delta a} = \boldsymbol{a'}-\boldsymbol{a''}$
separates the joint probability density into a quasiprobability signal ($Q$)
and detector noise ($D$).
\begin{eqnarray}
 && P(\boldsymbol{q}) =
    \int \rmd(\boldsymbol{\bar a}) \;
      D(\boldsymbol{q} - \boldsymbol{g}\cdot\boldsymbol{\bar{a}}) \,
      Q(\boldsymbol{\bar{a}}),\nonumber\\
  &&D(\boldsymbol{q} - \boldsymbol{g}\cdot\boldsymbol{\bar{a}}) = 
    \prod_k |\phi(q_k-g_k\bar{a}_k)|^2,\label{eq:weak_measurements_quasiprobability}\\
 && Q(\boldsymbol{\bar{a}}) =
    \int \rmd\boldsymbol{\delta a } \;
      \rme^{-(\boldsymbol{g}\cdot\boldsymbol{\delta a})^2/2} \;
          \tilde{\rho}_n
          (\boldsymbol{\bar{a}},\boldsymbol{\delta a} )\delta(\delta a_n).\nonumber  
\end{eqnarray}
Equation\ (\ref{eq:weak_measurements_quasiprobability}) defined the joint
quasiprobability density $Q$ for the series of von Neumann measurements. The quasiprobability has a well-defined 
limit $g\to 0$. In this limit for time-resolved measurement, the averages with respect to this quasiprobability are given by 
\begin{equation}
\langle a_1\cdots a_n\rangle=\int \rmd\boldsymbol a'\rmd\boldsymbol a''\,\delta(a'_n-a''_n)\tilde{\rho}(\boldsymbol a',\boldsymbol a'')\prod_k\frac{a'_k+a''_k}{2},
\end{equation}
which is equivalent to (\ref{tinor}).
The genuine, measured probability $P=Q\ast D$ is positive definite because it contains also the large detection noise $\sim 1$ which
is Gaussian, white and completely independent of the system, compared to the signal $\sim g$.

An alternative, equivalent approach is based on Gaussian positive operator-valued measures (POVMs) and special Kraus operators
\cite{povm,kraus,gpovm}. Let us begin with the
basic properties of POVM. The Kraus operators $\hat{K}(a)$ for an
observable described by $\hat{A}$ with continuous outcome $a$ need
only satisfy $\int \rmd a\hat{K}^\dag(a)\hat{K}(a)=\hat 1$. The act of
measurement on the state defined by the density matrix $\hat{\rho}$
results in the new state $\hat{\rho}(a) =
\hat{K}(a)\hat{\rho}\hat{K}^\dag(a)$. The new state yields a
normalized and positive definite probability density $P(a)=\mathrm{Tr}\,\hat{\rho}(a)$.  The
procedure can be repeated recursively for an arbitrary sequence of (not
necessarily commuting) operators $\hat{A}_1,\dots,\hat{A}_n$,
\begin{equation}
\hat{\rho}(a_1,\dots,a_n)=\hat{K}(a_n)\hat{\rho}(a_1,\dots,a_{n-1})
\hat{K}^\dag(a_n)\;.
\end{equation}
The corresponding probability density is given by
$P(a_1,\dots,a_n)=\mathrm{Tr}\;\hat{\rho}(a_1,\dots,a_n)$.
We now define a family of Kraus operators,
namely $\hat{K}_g(a) =
(g^2/2\pi)^{1/4}\exp(-g^2(\hat{A}-a)^2/4)$.  It is clear that
$g\to\infty$ should correspond to exact, strong, projective
measurement, while $g\to 0$ is a weak measurement and gives a
large error.
In fact, these Kraus operators are exactly those associated with
the von Neumann measurements previously described.
We also see that strong projection changes the state (by
collapse), while $g\to 0$ gives $\hat{\rho}(a)\sim \hat{\rho}$,
and hence this case corresponds to weak measurement.  However, the
repetition of the same measurement $k$ times effectively means one
measurement with $g\to kg$ so, with $k\to\infty$, even a
weak coupling $g\ll 1$ results in a strong measurement.  For an
arbitrary sequence of measurements, we can  write the final
density matrix as the convolution
\begin{equation}
  \hat{\rho}_{\boldsymbol g}(\boldsymbol a) = 
  \int \rmd a'\,\hat{\varrho}_{\boldsymbol g}(\boldsymbol a')
  \prod_k d_k(a_k-a'_k)
\label{conv}
\end{equation}
with $d_k(a)= \rme^{-g^2 a^2/2}\sqrt{g_k^2/2\pi}$.  Here
$\boldsymbol g=(g_1,\dots,g_n)$ $\boldsymbol
a=(a_1,\dots,a_n)$, and $\rmd a=\rmd a_1\dots \rmd a_n$. The quasi-density matrix
$\hat{\varrho}$ is given recursively by
\begin{eqnarray}
  \hat{\varrho}_{\boldsymbol g}(\boldsymbol a)  & = & 
  \int \frac{\rmd\xi}{2\pi}\rme^{-\rmi\xi a_n}
  \int\frac{\rmd\phi}{\sqrt{\pi g^2_n/2}}\rme^{-2\phi^2/g^2_n}
\label{quasi}\\
  &&
  \times\rme^{\rmi(\xi/2+\phi)\hat{A}_n}\hat{\varrho}_{\boldsymbol g}(a_1,\dots,a_{n-1})\rme^{\rmi(\xi/2-\phi)\hat{A}_n}\nonumber
\end{eqnarray}
with the initial density matrix $\hat{\varrho}=\hat{\rho}$ for $n=0$.
We can interpret $d$ in (\ref{conv}) as some internal noise of the detectors
which, in the limit $g\to 0$, should not influence the
system. We \emph{define} the quasiprobability \cite{ours} $Q_{\boldsymbol
  g}=\mathrm{Tr}\,\hat{\varrho}_{\boldsymbol g}$ and
abbreviate $Q\equiv Q_{\boldsymbol 0}$. In this limit
(\ref{quasi}) reduces to
\begin{equation}
\hat{\varrho}(\boldsymbol a)=\int\frac{\rmd\xi}{2\pi}\rme^{-\rmi\xi a_n}
\rme^{\rmi\xi\hat{A}_n/2}\hat{\varrho}(a_1,\dots,a_{n-1})\rme^{\rmi\xi\hat{A}_n/2}\,.
\label{gene}
\end{equation}
Note that $Q_{0\dots 0,g}=Q$, so the last
measurement does not need to be weak (it can be even a projection). 
The averages with
respect to $Q$ are easily calculated by means of the generating
function (\ref{gene}), e.g. $\langle
a\rangle_Q=\mathrm{Tr}\,\hat{A}\hat{\rho}$, $\langle
ab\rangle_Q=\mathrm{Tr}\,\{\hat{A},\hat{B}\}\hat{\rho}/2$,
$\langle
abc\rangle_Q=\mathrm{Tr}\,\hat{C}\{\hat{B},\{\hat{A},\hat{\rho}\}\}/4$
for $\boldsymbol a=(a,b,c)$. As a straightforward generalization to continuous measurement,
we obtain
\begin{eqnarray}
&&\langle a_1(t_1)\cdots a_n(t_n)\rangle_Q=\label{vrho}\\
&&\mathrm{Tr}\hat{\rho}\{\hat{A}_1(t_1),\{\cdots\{\hat{A}_{n-1}(t_{n-1}),
\hat{A}_n(t_n)\}\cdots\}\}/2^{n-1}\nonumber
\end{eqnarray}
for time ordered observables, $t_1\leq t_2\leq\cdots\leq t_n$.

\section*{Appendix B} 
\renewcommand{\theequation}{B.\arabic{equation}}
\setcounter{equation}{0}
\begin{figure}
\includegraphics[scale=.5]{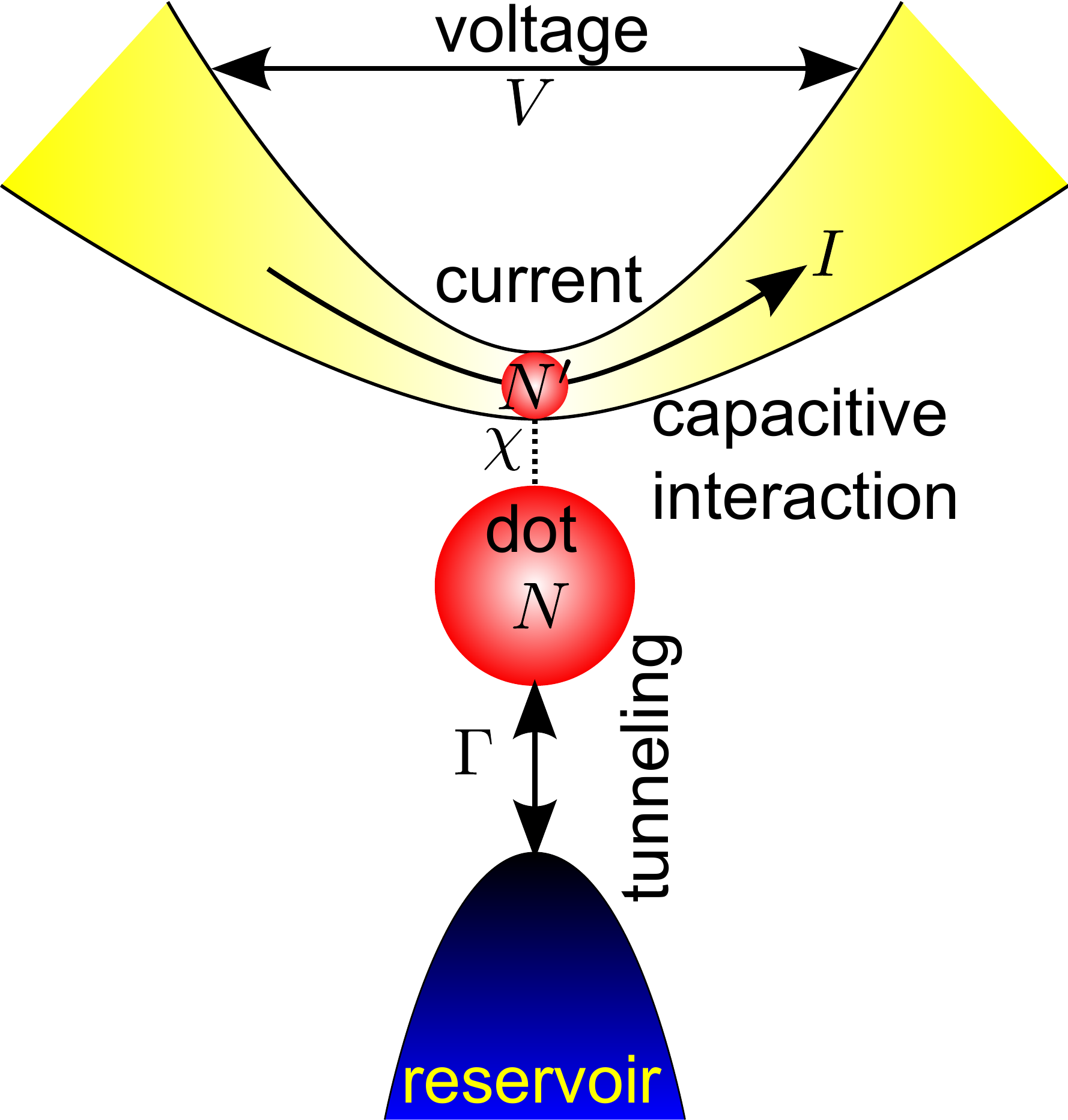}
\caption{The model of detecting the dot's charge. An electric junction contains another dot with effective occupation $N'$, coupled
capacitively to the measured dot. The fluctuations of the current $I$ in the junction biased 
by the voltage $V$ depend on the dot's occupation $N$ with the proportionality constant $\chi$.}
\label{qpc}
\end{figure}

An effective model of weakly detecting the dot's charge using an electric junction is shown in
Fig.\ \ref{qpc}. The junction is treated as another dot
between two reservoirs but
in a broad level regime. The complete Hamiltonian,
consisting of the dot part (\ref{hdot}), and the junction part, reads
\begin{eqnarray}
&&\hat{H}+\varepsilon'\hat{N}'+\hat{H}_V+e^2\hat{N}\hat{N}'/C+eV\hat{Q}_L
+\int \rmd E\;\times\nonumber\\
&&
\sum_{A=L,R} [\sqrt{\Gamma'/2\pi}\hat{d}^{\dag}\hat{\psi}_A(E)
+\mbox{h.c.}+E\hat{\psi}_A(E)\hat{\psi}_A(E)],\nonumber\\
&&\hat{N}_L=\hat{\psi}_L^{\dag}(E)\hat{\psi}_L(E),\:\hat{N}'=\hat{d}^{\dag}\hat{d},
\end{eqnarray}
where $\hat{N}_L$ is the total number of elementary charges $e$ in the left reservoir, $C$ is the capacitance between the dot and the QPC,
 $\Gamma'$,$\varepsilon'$ denote effective tunneling rate and level energy of the QPC and $V$ is the bias voltage.

We measure current fluctuations in the junction, $I(t)$,
with the current in Heisenberg picture defined as 
$\hat{I}(t)=-e\rmd\hat{N}_L(t)/\rmd t$.
Such fluctuations have already been measured experimentally at low
and high frequencies \cite{reulet-reznikov}.
Most of fluctuations are just generated by the shot noise in the junction.
Now, we consider a finite, but still very large capacitance.
We expect a contribution from the system dot's charge fluctuation to $S^{I}_3$ of the order
$C^{-3}$.
We assume separation
of the system's and detector's characteristic frequency scales, namely
\begin{equation}
(\Gamma,\varepsilon,k_BT)\ll eV\ll (\Gamma',\varepsilon'),
\end{equation}
which also includes the broad level approximation for the detector's dot. There exists a special parameter range,
\begin{equation}
\frac{e^2}{\Gamma'C}\gg\frac{\Gamma}{eV}\gg\left(\frac{e^2}{\Gamma'C}\right)^2,
\end{equation}
where the coupling is strong enough to extract information about $N(t)$ which
is not blurred by feedback and cross-correlation terms (left inequality),
but weak enough not to drive the system dot out of equilibrium (right inequality). In this limit the dominating contributions
to the detector current's third cumulant are given by 
$S_3^I\simeq S_3^{I0}+\chi^3S_3^N$
 with
\begin{eqnarray}
&&S_3^{I0}=\mathcal T(1-\mathcal T)(1-2\mathcal T)e^4V/h,\nonumber\\
&&
\chi=-e^2\rmd\langle I\rangle/C\rmd\varepsilon',
\end{eqnarray}
where $\langle I\rangle=\mathcal Te^2V/h$ and effective transmission
$\mathcal T=\Gamma^{\prime 2}/(\varepsilon^{\prime 2}+\Gamma^{\prime 2})$. Although the $\sim\chi^3$ term in $S^I_3$
is much smaller than the first one, other terms, corresponding to
cross correlations and back action, are negligible compared to the last term.



\end{document}